\input{aipcheck.tex}
  \def\selectedoptions{final}
\documentclass[
   \selectedoptions
  ]
  {aipproc}
\def\selectedlayoutstyle{6x9}
\layoutstyle\selectedlayoutstyle

\newcommand{\gray}{$\gamma$-ray}
\newcommand{\grays}{$\gamma$-rays}

\newcommand{\pubjournal}[6] {#1, \emph{#2}, {\bf #3}, #4 (#5)}
\newcommand{\pubjournala}[6]{#1, \emph{#2}, #4 (#5)}

\newcommand{\pubprocc}[6]{#1, ``#6,'' in \emph{#2}, #3, #5, p.#4}
\newcommand{\aap}{A\&A}
\newcommand{\adv}{Adv.\,Spa.\,Res.}
\newcommand{\apj}{ApJ}
\newcommand{\app}{Astropart.\,Phys.}

\newcommand{\icrc}{Int.\,Cosmic\,Ray\,Conf.}

\newcommand{\jgr}{J.\,Geoph.\,Res.}
\newcommand{\mnras}{Mon.\,Not.\,Roy.\,Astron.\,Soc.}
\newcommand{\nat}{Nature}
\newcommand{\physrep}{Phys.\,Rep.}

\newcommand{\prl}{Phys.\,Rev.\,Lett.}
\newcommand{\rpp}{Rep.\,Progr.\,Phys.}

\begin{document}
\title 
      [Diffuse gamma-ray emission]
      {Diffuse \gray\ emission: lessons and perspectives\vspace{-0.2em}}


\author{Igor V. Moskalenko}{
  address={Astroparticle Physics Laboratory, 
  NASA Goddard Space Flight Center, Greenbelt, MD 20771},
  altaddress={New address: Hansen Experimental Physics Laboratory, 
  Stanford University, Stanford, CA 94305}
}

\author{Andrew W. Strong}{
  address={Max-Planck-Institut f\"ur Extraterrestrische Physik, 
  Postfach 1312, 85741 Garching, Germany}
}

\copyrightyear  {2001}

\begin{abstract}
   The Galactic diffuse emission is potentially able to reveal much
about the sources and propagation of cosmic rays (CR), their spectra
and intensities in distant locations. It can possibly unveil WIMP dark
matter (DM) through its annihilation signatures. The extragalactic
background may provide vital information about the early stages of the
universe, neutralino annihilation, and unresolved sources (blazars?)
and their cosmological evolution. The \gray\ instrument EGRET on the
CGRO contributed much to the exploration of the Galactic diffuse
emission. The new NASA Gamma-ray Large Area Space Telescope (GLAST) is
scheduled for launch in 2007; study of the diffuse \gray\ emission is
one of the priority goals.  We describe current understanding
of the diffuse emission and its potential for future discoveries.
\end{abstract}
\vspace{-0.8em}

\date{\today}

\maketitle

\renewcommand{\baselinestretch}{0.95}

\section{Introduction}\vspace{-0.1em}
   Diffuse galactic emission dominates the \gray\ sky.  About
90\% of the total luminosity of the Milky Way galaxy at high energies
comes from the processes in the interstellar medium (ISM).  What
can we learn from studies of the diffuse emission?  First of all,
the diffuse emission is  a tracer of energetic interactions of
particles in the ISM and is produced via inverse Compton (IC)
scattering, bremsstrahlung, and $\pi^0$-decay.  It thus delivers
information about spectra and intensities of CR species, primarily
protons, He, electrons, and positrons, at distant locations and allows
one to study CR acceleration in supernova remnants (SNRs) and
other sources as well as propagation in the ISM. Emission from the local
ISM can be used to study the local CR spectrum and thus provides a
valuable input for studies of solar modulation and CR variations in the
Galaxy.
\grays\ can be used to trace interstellar gas independently of other
astronomical methods, e.g.\ the relation of molecular H$_2$ gas to
the CO molecule \cite{S04} and hydrogen overlooked by other methods 
\cite{grenier05}.
The diffuse emission may contain signatures of new physics
(e.g., DM) or may be used to put restrictions on cosmological models
and the parameter space of supersymmetric particle models,  and/or
provide some hints for accelerator experiments. On the other hand,
the diffuse emission is a background for point sources and its
accurate determination is important for accurate localization of such
sources and their spectra especially at low latitudes.  It is also a
foreground in studies of extragalactic diffuse emission, which is in
itself is a complicated subject and may contain information about
early stages of the universe and new physics (DM, black hole
evaporation etc.).  Besides, GLAST may be able to detect and spatially
resolve the \emph{diffuse} \grays\ from normal galaxies
and thus will enable us to study CR in other galaxies \cite{digel00}.
This will give us an external view of a sister spiral galaxy
Andromeda (M31) much like the Milky Way, and the dwarf irregular 
galaxies LMC and SMC.

\renewcommand{\baselinestretch}{1}

\section{Diffuse gamma rays and cosmic rays}
   To understand the diffuse emission one needs to know particle
spectra in the entire Galaxy.  The conventional sources of CR are
believed to be supernovae (SNe) and SNRs, pulsars, compact objects in
close binary systems, and stellar winds.  Observations of synchrotron
emission (e.g., \cite{koyama95}),
\grays\ \cite{thompson00}, and even TeV \grays\ \cite{aharonian1} 
reveal the presence of energetic particles in these objects thus
testifying to efficient acceleration processes.  Propagation in the
ISM changes the initial composition and spectra of CR species due to
the energy losses (ionization, Coulomb scattering, bremsstrahlung, IC
scattering, and synchrotron emission), energy gain (diffusive
reacceleration), and other processes (i.e., diffusion and convection
by the Galactic wind).  The destruction of primary nuclei via
spallation gives rise to  secondary nuclei and isotopes which are rare
in nature (i.e., Li, Be, B), antiprotons, and pions ($\pi^\pm$,
$\pi^0$) that decay producing secondary $e^\pm$'s and \grays.
\grays\ are also produced by electrons (and positrons) 
via IC and bremsstrahlung.  The observation of diffuse \grays\ gives
us the intensity integrated over the line of sight and provides the
most direct probe of the proton and electron spectra in distant
locations.

The propagation parameters are determined from studies of the nuclear
component in CR using the transport equation
\cite{berezinskii90}, which may also include 
convection by a Galactic wind (e.g.,
\cite{Zirakashvili96}), distributed acceleration in the ISM due to the
Fermi  2nd-order mechanism \cite{seo94} (so-called ``reacceleration''),
and non-linear wave-particle interactions \cite{ptuskin03}.  The
abundances of stable (Li, Be, B,  Sc, Ti, V) and radioactive
($^{10}$Be, $^{26}$Al, $^{36}$Cl, $^{54}$Mn) secondaries in CR are
used to derive the diffusion  coefficient and the halo size
\cite{SM98,ptuskin-webber98}.

K-capture isotopes in CR  (e.g., $^{49}$V, $^{51}$Cr) can  serve as
important energy markers and can be used to study the energy-dependent
effects such as diffusive  reacceleration in the ISM and heliospheric
modulation \cite{soutoul98}.  Such nuclei usually decay via
electron-capture and have a short lifetime in the medium; in CR they
are stable or live longer as they are created bare by fragmentation of
heavier nuclei.  At low energies, their lifetime depends on the
balance between the energy-dependent probabilities of the electron
attachment from the ISM and stripping.

Study of the light nuclei in CR (Li--O) allows us to determine
propagation parameters averaged over a larger Galactic region, but the
local ISM is \emph{not} necessarily the same and the \emph{local}
propagation  parameters may significantly differ. To probe the local
ISM one should look at isotopes with shorter lifetimes (e.g.,
$^{14}$C) \cite{yanasak01} 
and heavy nuclei since larger fragmentation cross sections
lead to a smaller ``collection area.''  The CR source composition is
derived from direct CR data by correcting for the effects of
propagation, spallation, and solar modulation. The derived source
abundances may provide some clues to mechanisms and sites of CR
acceleration.  Obviously, such information is also valuable for a
number of astrophysical/astroparticle applications far beyond the
scope of the diffuse emission. For a more comprehensive review see
\cite{M04rev,M05rev}.

\subsection{The GeV excess}
The puzzling excess in the EGRET diffuse emission data above 1 GeV
relative to that expected \cite{hunter97,SMR00} has shown up in all
models that are tuned to be consistent with directly measured CR
nucleon and electron spectra \cite{M04rev,SMR04a}. 
The excess has shown up in all directions, not only in the Galactic plane;
it is possible but unlikely to be an instrumental 
artefact due to the uncertainty in calibration.

The CR spectrum in the local ISM should be similar to the one
directly measured corrected for the effects of solar modulation;
thus future observations of \gray\ emission from
local ISM could provide some clues to its origin.
If the GeV excess appears in the emission from the local ISM, it may
be either the result of the poor knowledge of pion production cross
section (e.g., \cite{kamae05}) or the DM signal.
If the local ISM does not exhibit the GeV excess seen on the large
Galactic scale, it
probably has something to do with CR spatial fluctuations. In the
latter case, the normalization  of the global calculated CR spectrum
to the local CR measurements is wrong since the local spectrum is not
representative of the \emph{local} Galactic average and we should look
for another way of fixing the global normalization.  Currently
available EGRET data on the local clouds  have large error bars (see
\cite{digel01} for a summary plot and references) and therefore are
not conclusive.

Secondary $\bar p$'s are produced in the same interactions  of CR
particles with interstellar gas as diffuse \grays\ and $e^+$'s and
provide another probe of the interstellar CR proton spectrum.
Recent $\bar p$ data with larger statistics
\cite{Orito00-Maeno01-beach-boezio01-bergstrom} 
triggered a series of calculations of the secondary $\bar p$  flux in
CR. The observed flux has been shown to be consistent with  their
secondary origin in calculations using plain diffusion (or the
leaky-box model) and a Galactic wind model \cite{simon98,M02}.  The
diffusive reacceleration models have certain  advantages compared to
other propagation models:  they naturally reproduce secondary/primary
nuclei ratios in CR and have only three free parameters.  Detailed
analysis shows, however, that the reacceleration models
\emph{underproduce} $\bar p$'s by a factor of $\sim$2 at 2 GeV \cite{M02}
where the statistics are largest; this is because matching the B/C
ratio at all energies requires the diffusion coefficient to be too
large. Note that this is an essential feature of reacceleration
models.
\vspace{-0.5em}

\section{Hypotheses}
   There are quite a few hypotheses trying to explain the \gray\ and
$\bar p$ excesses.
Among these are: harder CR proton \cite{mori97} and/or electron
spectra \cite{SMR00,porter97} in distant Galactic regions,  a
local CR component at low energies perhaps associated with the Local
Bubble \cite{M03}, contribution of unresolved point sources
\cite{wang05} or SNRs with freshly accelerated particles
\cite{voelk04}, CR spatial intensity variations \cite{SMR04a},  and
DM signals (e.g., recent discussions
\cite{cesarini04,deboer05}).

The harder CR proton spectrum hypothesis has been shown to be problematic
in the past \cite{M98} because
the harder proton spectrum needed to explain diffuse \grays\
would produce too many $\bar p$'s.
The harder electron spectrum model is
more feasible, but does not resolve the $\bar p$ excess.
The contribution of
unresolved  point sources and SNRs may explain some part of the \gray\ GeV
excess in the Galactic plane, but not at high Galactic latitudes.  A
low energy local component in CR, while improving agreement with
$\bar p$'s does not explain the excess in \grays.

As will be explained in the following sections, we
are, in fact, left with at least two possibilities to explain the excesses
in \grays, and $\bar p$'s, simultaneously: to invoke
CR intensity variations or DM signals.

\subsection{CR intensity variations}
   There are a number of reasons why CR intensity may fluctuate in
space and time. First is the stochastic nature of SN events.
Dramatic increases in CR intensity perhaps connected
with SN explosions nearby are recorded in terrestrial concentrations
of  cosmogenic isotopes.  Concentrations of $^{10}$Be in Antarctic and
Greenland ice core samples
\cite{yiou97} and $^{60}$Fe in a deep-sea
ferromanganese crust \cite{knie04} indicate highly significant
increases of CR intensity $\sim$40 kyr and 2.8 Myr ago,
correspondingly. The SN rate is larger in the spiral arms
\cite{case96}; this may lead to lower CR intensity in the interarm
region where the solar system is located. In case of anisotropic
diffusion or convection by the Galactic wind such fluctuations may be
even stronger.  
The intensity variations and spectra of CR protons
and  heavier CR nuclei may be uncorrelated.  The total inelastic cross
section for protons is $\sim$30 mb vs.\ $\sim$300 mb for Carbon,
so that their Galactic ``collecting areas'' 
differ by a factor of 10 or more
for heavier nuclei; this possibly implies that  the directly measured
CR protons and CR nuclei come from different sources.  Besides, CR
electrons and positrons suffer large energy losses
\cite{SM98} and thus their spectral and intensity fluctuations
can be considerable.

Antiprotons in CR are presumably secondary and produced by mostly  CR
protons in interactions with the interstellar gas.  Because of their
secondary origin, their intensity fluctuates less than that of
protons.  The total cross section of $\bar p$'s is about the same as
protons, except at low energies due to the annihilation, and thus they
trace the \emph{average} proton spectrum in the Galaxy.
If the directly measured
local CR spectrum is not representative of the \emph{local} Galactic
average, the
\emph{antiproton measurements} can still be used instead
to derive the intensity of CR \emph{protons}.

When normalized to the local CR proton spectrum, the reacceleration
model underproduces $\bar p$'s and diffuse \grays\ above 1 GeV 
by the \emph{same} factor of $\sim$2 \cite{SMR00,M02} 
while it works well for other CR
nuclei. It is thus enough to renormalize the CR  proton spectrum up by
a factor of 1.8 to remove the excesses; the model then predicts  a
factor of 2 too many photons at $\sim$100 MeV. The 100 MeV photons are
produced mostly by $\sim$1 GeV protons where many uncertainties
simultaneously come into play: poor knowledge of the
$\pi^0$-production cross section at \emph{low} energies and/or
low-energy interstellar proton spectrum and/or solar modulation (see
\cite{M03} for more discussion).  To get agreement with the EGRET
data effectively requires a corresponding adjustment of the
spectrum of CR protons at low energies. Since the IC and $\pi^0$-decay
photons have different distributions, the electron spectrum also needs
to be renormalized up by a factor of 4 using the diffuse \gray\  flux
itself. Thus, our ``optimized'' model is able to reproduce 
the spectrum of the diffuse
\grays\ in \emph{all directions} as well as the
latitude and  longitude profiles for the whole EGRET energy range 30
MeV -- 50 GeV, and is consistent with our current understanding of CR
variations in the Galaxy \cite{SMR04a}.

A logical extension of this analysis is a new determination of the
extragalactic background \cite{SMR04b}, which is lower than previously
thought due to the increased Galactic component, mostly IC.  It has
also a positive curvature as expected in blazar population studies and
various scenarios of cosmological neutralino annihilation
\cite{stecker96-mannheim}.

\subsection{Dark Matter}
   The existence of non-luminous DM is now generally accepted by
most of the astrophysical community though its nature is uncertain.
The major division is between baryonic and non-baryonic DM with the
preference given to the non-baryonic option over the last years.  A
number of particle DM candidates is discussed in the literature
\cite{jkg96-bergstrom00-bertone05}, where the most popular are
the lightest supersymmetric neutralino $\chi^0$ and a Kaluza-Klein
hypercharge $B^1$ gauge boson.  Annihilation of DM particles creates a
soup of standard and  supersymmetric constituents, which eventually
decays to ordinary  mesons, baryons, and leptons.  The DM particles in
the halo or at the Galactic center  may thus be detectable via their
annihilation products  ($e^+$, $\bar p$, $\bar d$, \grays) in CR.  The
standard approach is to scan the SUSY parameter space
\cite{darksusy05} to find a candidate able to fill the excesses in
diffuse \grays, and/or $\bar p$'s, and/or $e^+$'s over the predictions
of a conventional model (as discussed in the previous sections).  The
diffuse \gray\ data are far richer  than the CR particle data because
\grays\ are coming from distant  locations and different directions on
the sky.  The EGRET data used in a recent analysis \cite{deboer05}
reveal a ``perturbed'' DM halo profile of the Milky Way possibly
indicating galactic mergers in the past. Though very interesting,
these results may only be taken as a hint which will require a lot of
work to confirm. The DM search is one of the
primary goals of the near-future space missions and the new CERN collider
LHC.

\section{Conclusion}
   There are essential lessons which we have learned from EGRET.  The
diffuse \gray\ flux is consistent with being produced mostly in
energetic CR particle ($p$, He, $e^\pm$)  interactions with
interstellar gas and radiation fields. However, there is an excess
above $\sim$1 GeV observed in all directions in the sky.
Understanding the excess is important
because it contains possibly new astrophysical information.
The \gray\ data must be
evaluated in conjunction with CR  particle data because this may lead
to better understanding of both CR and diffuse \grays.

The near future prospects are encouraging: several missions
complementing each other are planned. The first one  will be PAMELA to
be launched in December of 2005 and designed to measure $\bar p$'s,
$e^\pm$'s, and isotopes H--C over 0.1--300 GeV.  Though its detector
is small, it will provide enough exposure  during its three-year
mission.  BESS-Polar instrument has had a successful test flight in
Antarctica in winter 2004--05, and is planned to launch  again in 2006
during the next solar minimum. It will provide accurate data on $\bar
p$'s and light elements.  The new \gray\ telescope, the GLAST mission,
is scheduled for launch in the fall of 2007. It will be capable  of
measuring $\gamma$-rays in the range 20 MeV -- 300 GeV with much
better sensitivity and resolution than EGRET;  studies of the diffuse
emission is one of its primary goals.  The AMS mission should make
its flight to the  International Space Station in 2008 and will
measure  CR particles and nuclei
$Z\hbox{\rlap{\hbox{\lower3pt\hbox{$\sim$}}}\lower-2pt\hbox{$<$}}26$
from GeV to TeV energies.

\smallskip
This work was supported in part by grants from NASA Astrophysics Theory 
Program (ATP) and NASA Astronomy and Physics Research and Analysis (APRA) 
Program.

\bibliographystyle{aipproc}

\end{document}